\documentclass[aps,pra,twocolumn,amsmath,nofootinbib]{revtex4-1}
\usepackage{booktabs}
\usepackage{threeparttable}
\usepackage{epsfig}
\usepackage[colorlinks,linkcolor=blue,anchorcolor=blue,
            citecolor=blue,urlcolor=blue,
            breaklinks=true]{hyperref}
\usepackage{graphics}
\usepackage{epstopdf}
\usepackage{color}
\usepackage{dcolumn}
\usepackage{bm}
\usepackage{cases}
\usepackage{appendix}
\usepackage{lineno,hyperref}
\usepackage{ulem}

\newcommand{\tb}{\textbf}

\begin{document}
\author{Run Cheng$^{1,2}$}
\author{Yong-Long Wang$^{2,3}$}
\email{wangyonglong@lyu.edu.cn}
\author{Hao-Xuan Gao$^{1,2}$}
\author{Hao Zhao$^{1,2}$}
\author{Jia-Qi Wang$^{1}$}
\author{Hong-Shi Zong$^{1,3,4}$}
\email{zonghs@nju.edu.cn}
\address{$^{1}$ Department of Physics, Nanjing University, Nanjing 210093, P. R. China}
\address{$^{2}$ School of Physics and Electronic Engineering, Linyi University, Linyi 276005, P. R. China}
\address{$^{3}$ Joint Center for Particle, Nuclear Physics and Cosmology, Nanjing 210093, P. R. China}
\address{$^{4}$ State Key Laboratory of Theoretical Physics, Institute of Theoretical Physics, CAS, Beijing 100190, P. R. China}

\title{\normalsize{Geometric effects on the electronic structure and the bound states in annular corrugated wires}}
\begin{abstract}
In the spirit of the thin-layer quantization scheme, we give the effective Hamiltonian describing the noninteracting electrons confined to an annular corrugated surface, and find that the geometrically induced potential is considerably influenced by corrugations. By using numerical calculation, we investigate the eigenenergies and the corresponding eigenstates, and find that the transition energies can be sufficiently improved by adding corrugations. Particularly, the transition energy between the adjacent eigenstates corresponds to energy levels difference based on the wavefunction of annular wire, and the number of the energy levels is equal to the number of corrugations. And the larger magnitude of corrugations is capable of increasing the number of bound states. In addition, the distribution of ground state probability density is reconstructed by the corrugations, and the energy shift is generated.

\bigskip

\end{abstract}
\maketitle

\section{INTRODUCTION}
The rapid development of nanotechnology has realized the nanomaterials with complex geometries, such as bent nanotubes~\cite{Goldstone1992,Ouyang1999}, corrugated carbon nanotubes~\cite{Shima2010,Novakovic2011,Moraes2016Geometric,condmat4010003}, rolled-up nanotubes~\cite{Ortix2010}, and M\"{o}bius nanostructures ~\cite{Tanda2002,Gravesen2005}. These experimental realizations trigger the need for understanding the quantum motion of a particle confined to a deformed surface. Among them, the surface with corrugations is an important kind of two-dimension(2D) systems. In this case, a curvature-induced potential~\cite{Costa1981,Jaffe2003Quantum} appears in the effective Hamiltonian. A number of papers have shown that the geometric potential can generate localized surface states~\cite{Goldstone1992,Cantele2000Topological,Aoki2001Electronic,Encinosa2003,Taira2007Curvature,Taira2007Electronic,Ortix2010Effect,DU201628,Exner2019a}, change the band structures~\cite{Aoki2001Electronic,Fujita2005Band, Aoki2005Electronic,Chen2009Design,condmat4010003} and prompt the energy shifts~\cite{Encinosa1998,Shima2009Geometry}.

Recently, theoretical and experimental investigations have shown that the nanomaterials with complex geometries exhibit novel mechanical properties and provide new ways to construct the nanodevices with required electronic structure~\cite{Aoki2001Electronic,Aoki2005Electronic,Jensen2009,Ortix2010Effect,Grivickas2019}. In order to learn clearly the geometric effects on the electronic properties, we will consider a quantum system, electrons without interactions confined to an annular corrugated surface. For the quantum confinement system, it can be taken as two one-dimensional(1D) quantum components, because that the component of annular corrugated wire can be separated from that of $z$-axis analytically. And the effective quantum dynamics of 1D systems with deformations have been discussed widely, such as reflectionless quantum wire~\cite{doi:10.1143/JPSJ.60.3640}, deformed quantum wires~\cite{CLARK1996, Cantele2000Topological, Zhang2007Quantum, Schindler2019, Grivickas2019}, and Si nanowire~\cite{Duan2003}. These known results inspire us to consider the presence of corrugations in the annular surface. We will investigate the geometric effects of corrugations, and how the electronic properties are related to the number and the magnitude of corrugations.

This paper is organized as follows. In Section II, the effective Hamiltonian describing noninteracting electrons confined to an annular corrugated surface is given directly. And how the geometric potential is affected by the corrugations is also discussed carefully. In Section III, the energy level structure and the corresponding eigenstates for the annular wire component of the confined electron are investigated in the presence of geometric potential. In Section IV, the energy shift of the ground state determined by the geometric potential and the reconstructed distribution of the ground state probability density are calculated. Finally, in Section V, conclusions are briefly given.

\section{Quantum Dynamics of a Particle Confined to an Annular Corrugated  Wire}
\begin{figure}[htbp]
\centering
\includegraphics[width=0.44\textwidth]{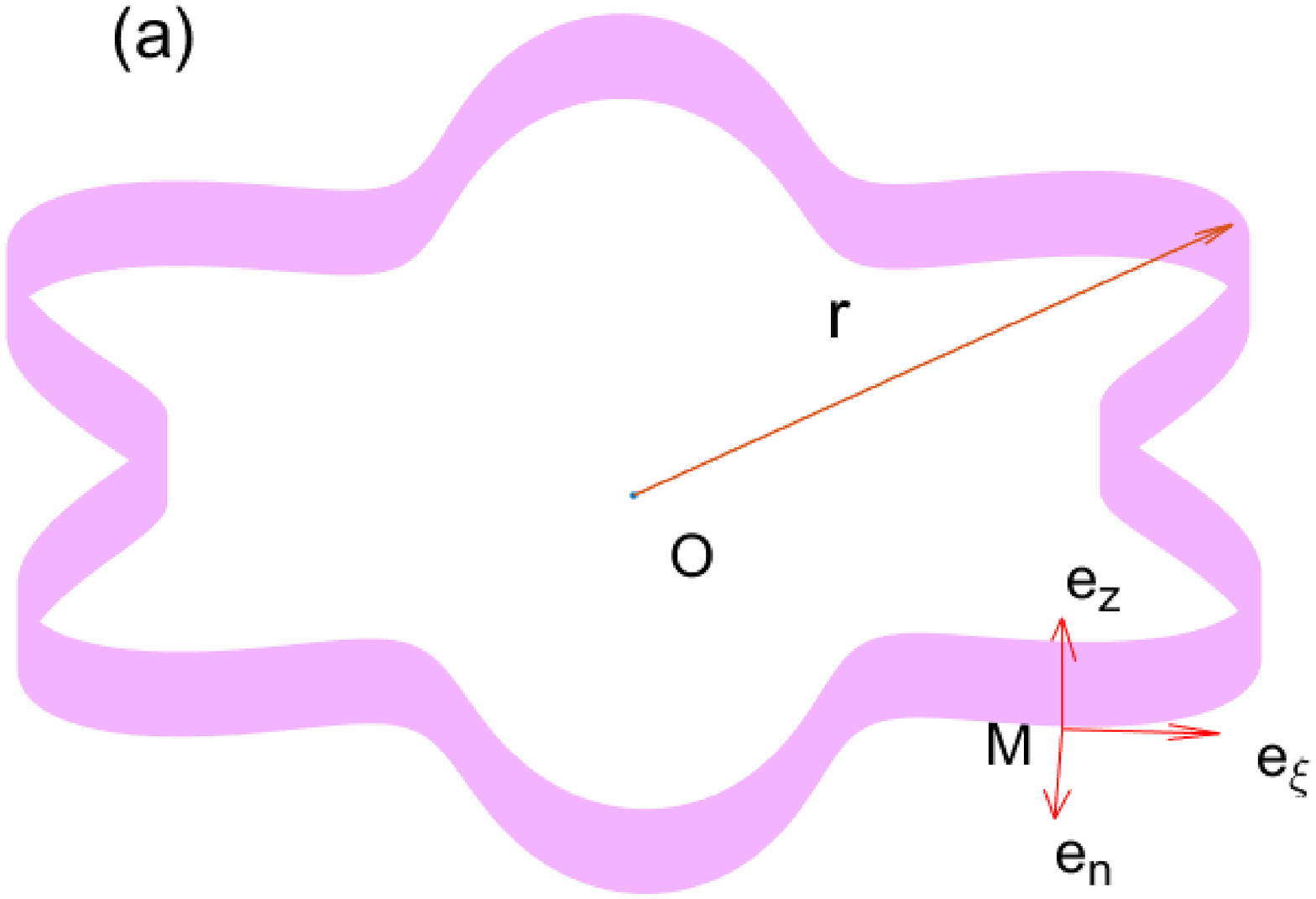}
\includegraphics[width=0.44\textwidth]{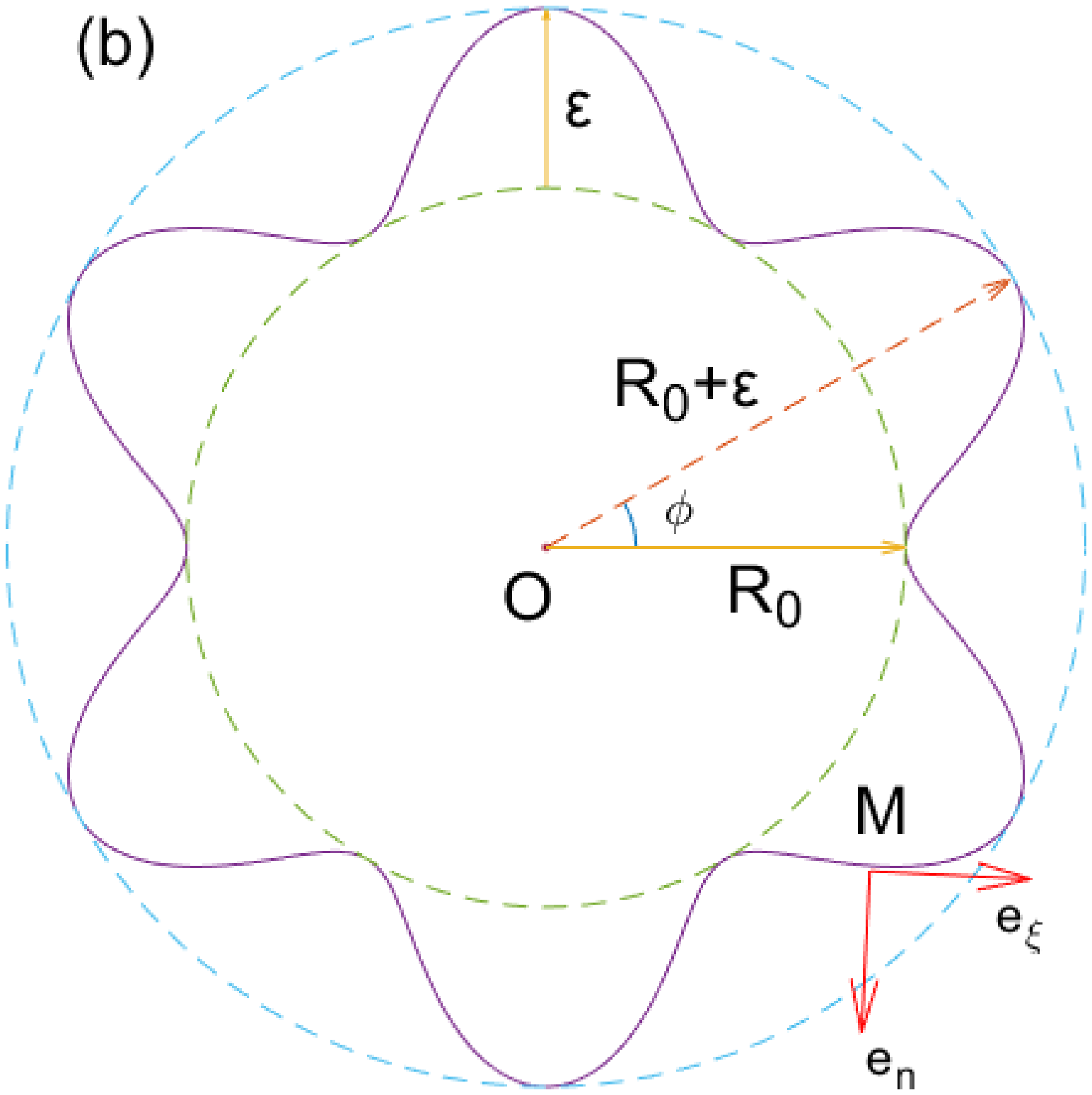}
\caption{\footnotesize (Color online) (a) Schematic of an annular corrugated surface described by $\tb{r}=(x,y,z)$ with $R_{0}$=4, $\epsilon$=2 and $N=6$. (b) Schematic of an annular corrugated wire described by $\tb{r}=(x,y)$ with $R_{0}$=4, $\epsilon$=2 and $N=6$. Here $R_{0}$ and $\phi$ are the radius of the undeformed part of the wire and the azimuthal angle, $\epsilon$ is the magnitude of corrugations.}\label{fig1}
\end{figure}

In this section, the effective Hamiltonian for an electron confined to an annular corrugated wire will be given in terms of the thin-layer quantization scheme~\cite{Jensen1971Quantum,Costa1981,Wang2017Geometric}. The study begins with an annular corrugated surface (see Fig.\ref{fig1}(a)) that is described by
\begin{equation}\label{f0}
\tb{r}=(x,y,z),
\end{equation}
where
\begin{equation}\label{f1}
\begin{split}
     & x=R\cos\phi, \\
     & y=R\sin\phi,\\
     & z=z,
\end{split}
\end{equation}
with
\begin{equation}
R=R_{0}+\frac{\epsilon}{2}[1-\cos(N\phi)],
\end{equation}
wherein ($R$, $\phi$) are the radius and azimuthal angle of the annular wire with corrugations sketched in Fig.\ref{fig1}(b), (x, y, z) describe the three coordinate variables of Cartesian coordinate system, $\epsilon$ and $N$ denote the magnitude and the number of corrugations, respectively.

According to Eq.~\eqref{f1}, by introducing the arc length of annular corrugated wire, $d\xi$=$\frac{1}{2}\sqrt{Q^2+S^2}$$d\phi$, two tangent unit basis vectors  $\tb{e}_{\xi}$, $\tb{e}_{z}$ and a normal unit basis vector $\tb{e}_{n}$ can be obtained as
\begin{equation}
\begin{split}
&\tb{e}_{\xi}=\frac{1}{U}(-Q\sin\phi+S\cos\phi,Q\cos\phi+S\sin\phi,0),\\
&\tb{e}_{z}=(0,0,1),\\
&\tb{e}_{n}=\frac{1}{U}(Q\cos\phi+S\sin\phi,Q\sin\phi-S\cos\phi,0),
\end{split}
\end{equation}
where
\begin{equation}\label{MyNote}
\begin{split}
& Q=2R_{0}+\epsilon-\epsilon\cos N\phi,\\
& S=\epsilon N\sin N\phi,\\
& U=\sqrt{Q^2+S^2}.
\end{split}
\end{equation}

\begin{figure}[htbp]
\centering
\includegraphics[width=0.46\textwidth]{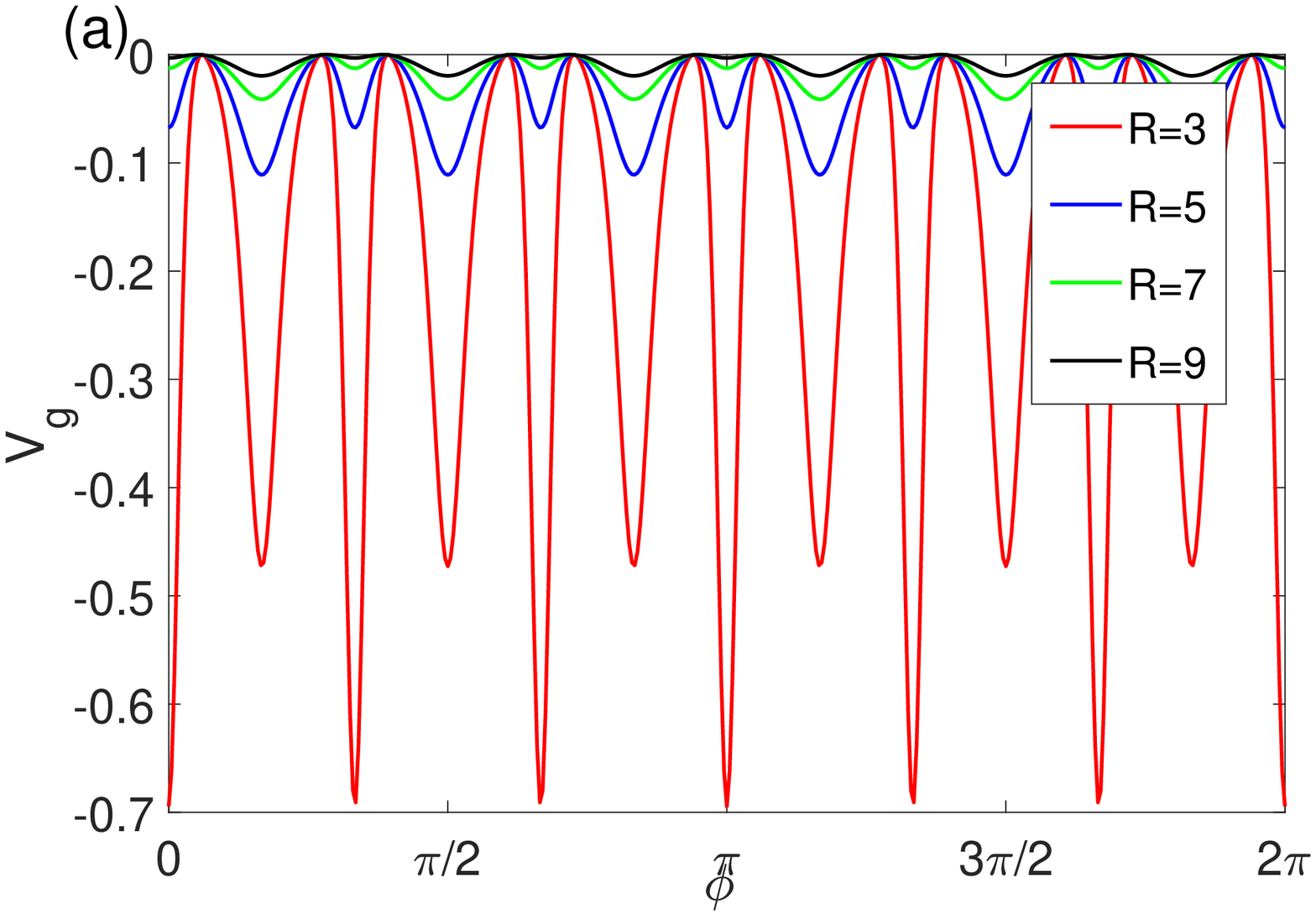}
\includegraphics[width=0.46\textwidth]{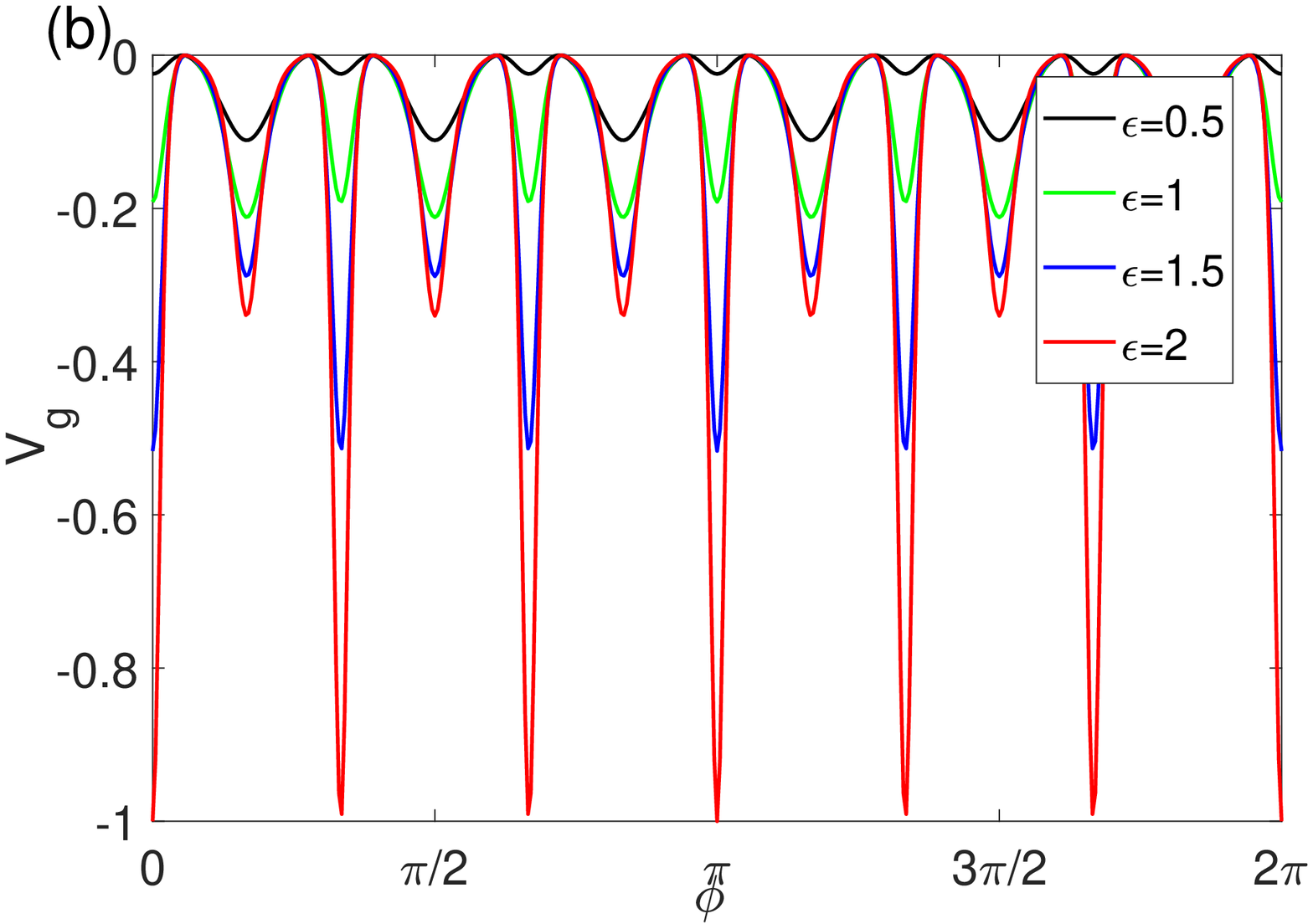}
\includegraphics[width=0.46\textwidth]{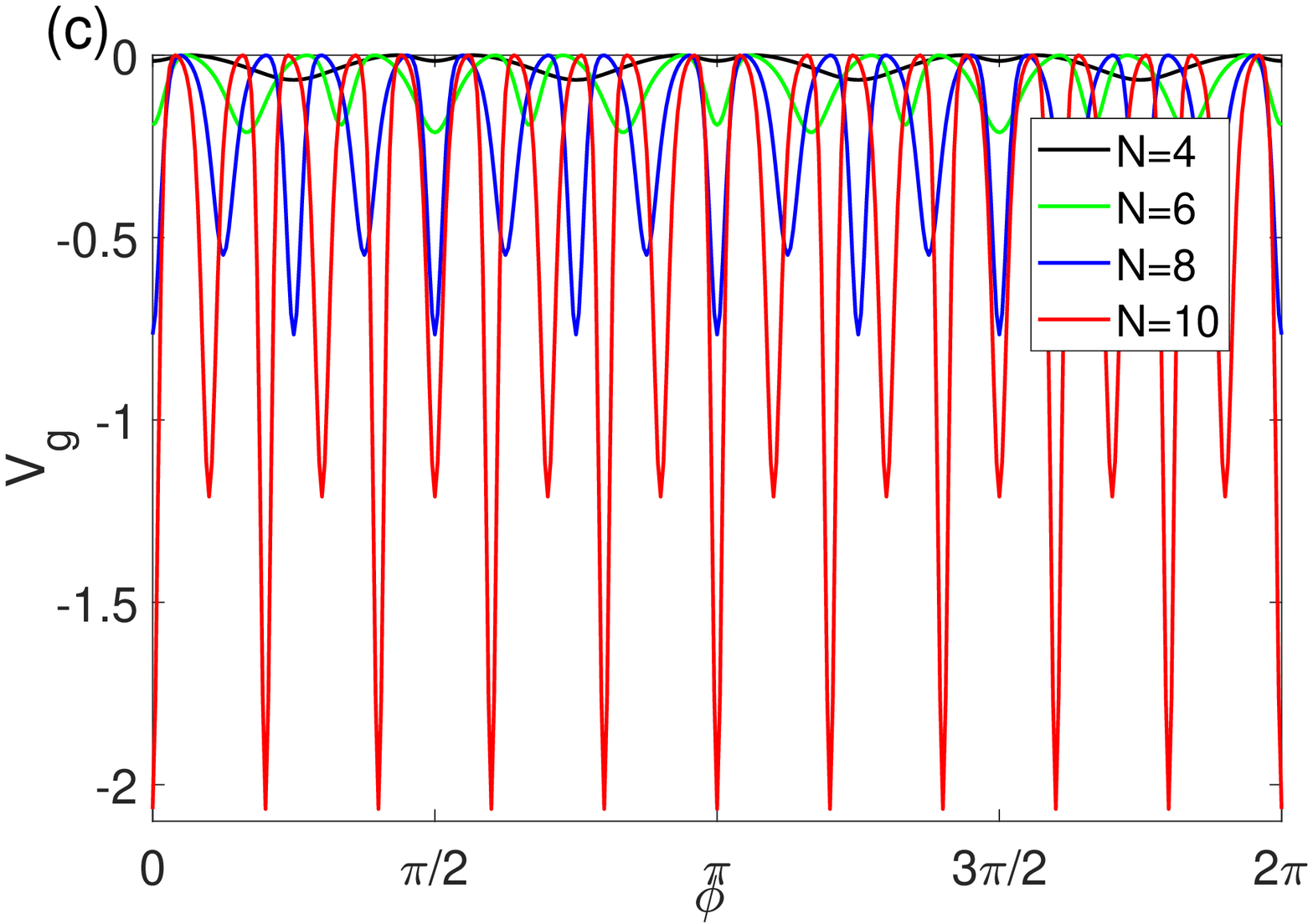}
\includegraphics[width=0.46\textwidth]{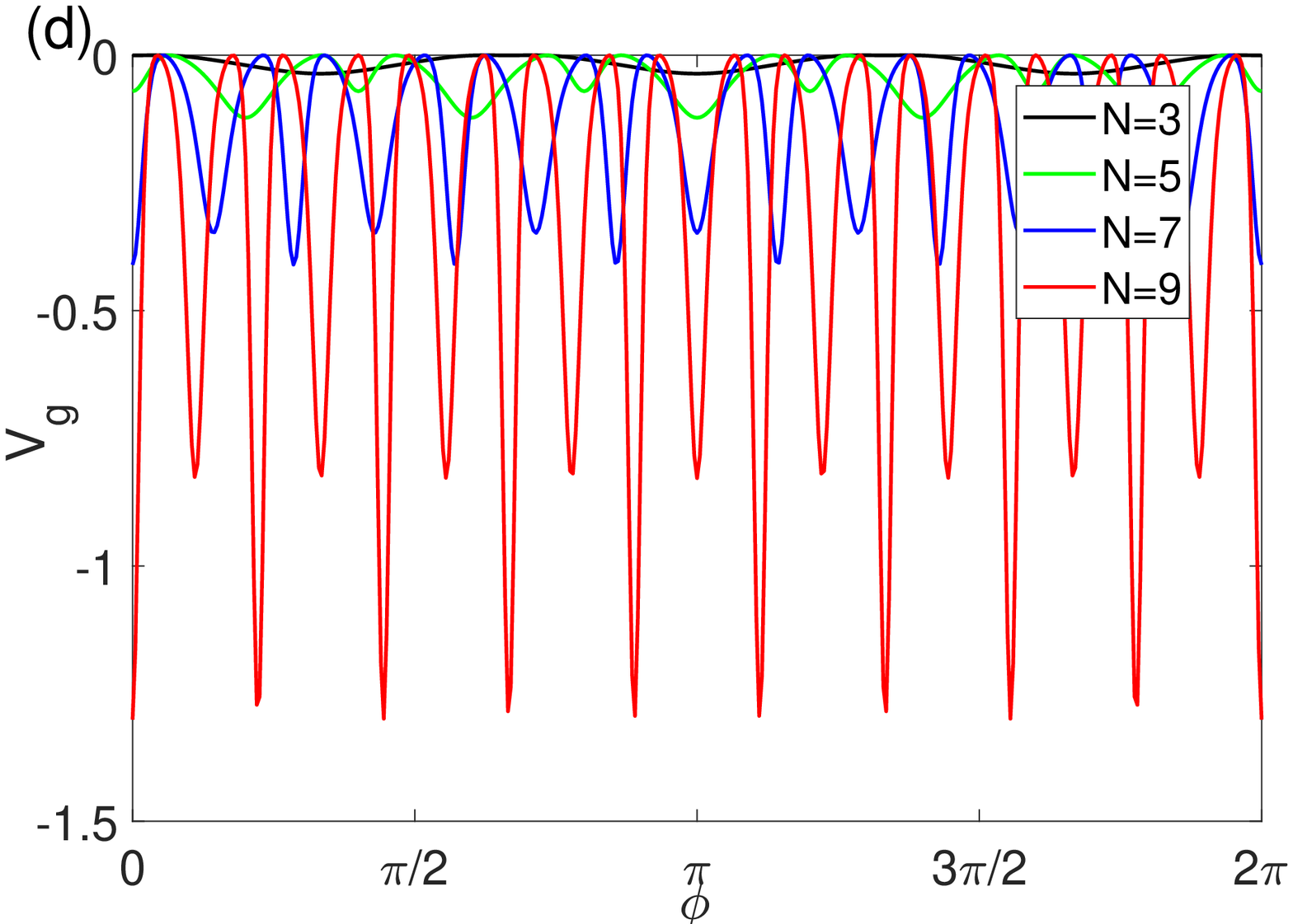}
\caption{\footnotesize (Color online) The geometric potential versus $\phi$ for the cases of (a) $N=6$, $\epsilon=1$ and $R_0=3, 5, 7, 9$, (b) $N=6$, $R_{0}=4$ and $\epsilon=0.5, 1, 1.5, 2$, (c) $R_{0}=4$, $\epsilon=1$ and $N=4, 6, 8, 10$, (d) $R_{0}=4$, $\epsilon=1$ and $N=3, 5, 7, 9$. Here $\frac{\hbar^2}{2m^*}$ is taken as an unit.}\label{corrugated GPv}
\end{figure}

And then the position vector $\tb{R}$ of a point close to the annular corrugated surface can be described by
\begin{equation}
\tb{R}=\tb{r}+q_3\tb{e}_n,
\end{equation}
where $q_3$ is the curvilinear coordinate variable along $\tb{e}_n$. In terms of the two position vectors $\tb{r}$ and $\tb{R}$, with the definitions
$g_{ab}=\partial_a\tb{r}\cdot\partial_b\tb{r}$ $(a, b=\xi, z)$ and $G_{ij}=\partial_i\tb{R}\cdot\partial_j\tb{R}$ $(i, j=\xi, z, n)$, the covariant components of surface metric tensors are
\begin{equation}\label{surfmetric}
g_{\xi\xi}=1, g_{zz}=1, g_{\xi z}=g_{z \xi}=0,
\end{equation}
and those metric tensors defined in the subspace as
\begin{equation}\label{spacemetric}
\begin{split}
& G_{\xi\xi}=(1-q_3 w)^2,\\
& G_{\xi z}=G_{z\xi}=G_{\xi n}=G_{n\xi}=0,\\
& G_{zz}=G_{nn}=1,G_{zn}=G_{nz}=0,
\end{split}
\end{equation}
with
\begin{equation}\label{eq1}
w=\frac{2(Q^2+2S^2-\epsilon N^2Q\cos N\phi)}{(Q^2+S^2)^{\frac{3}{2}}}.
\end{equation}
Obviously, $g$ and $G$ satisfy the relationship $G$=$f^2$$g$, where $g$ and $G$ are the determinants of $g_{ab}$ and $G_{ij}$, and $f$ is the rescaling factor, $f$=1-$q_3$$w$. The covariant components of the Weingarten curvature tensor are given by
\begin{equation}\label{weingarten curvature tensor}
\begin{split}
 & \alpha_{\xi\xi}=w, \\
 & \alpha_{\xi z}=\alpha_{z\xi}=\alpha_{zz}=0.
\end{split}
\end{equation}
The mean curvature is then given by $M=\alpha_{\xi\xi}/2$, and the Gaussian curvature $K$ is zero. Following the Ref.~\cite{Ferrari2008}, with the metric tensors~\eqref{surfmetric} and~\eqref{spacemetric}, the effective Hamiltonian describing an electron confined to the annular corrugated surface is obtained as
\begin{equation}\label{h1}
H_{2D}=-\frac{\hbar^2}{2m^*}\partial^2_{\xi}-\frac{\hbar^2}{2m^*}(M^2-K)-\frac{\hbar^2}{2m^*}\partial^2_{z},
\end{equation}
where $\hbar$ is the Plank constant divided by $2\pi$, $m^*$ is the effective mass of an electron. In the lights of the Hamiltonian ~\eqref{h1}, it is easy to separate the quantum motion in $z$-direction from the rest part. As ansatz, the surface wave function can be expressed as
\begin{equation}\label{surface function}
\psi(\xi,z)=\psi(\xi)\times e^{ik_{z}z},
\end{equation}
where $k_{z}$ is the $z$-component of momentum. And then the effective Hamiltonian of $\xi$-component can be written as
\begin{equation}\label{H1D}
H_{\rm{eff}}=K+V_g,
\end{equation}
where $K$ is the kinetic energy operator for an electron constrained to move along the annular corrugated wire,
\begin{equation}\label{kinetic}
\begin{split}
 &K=-\frac{\hbar^2}{2m^*}\frac{\partial^2}{\partial\xi^2}\\
 &\quad=-\frac{\hbar^2}{2m^*}\frac{4}{Q^2+S^2}[\frac{\partial^2}{\partial\phi^2}-\frac{S(\epsilon N^2\cos N\phi+Q)}{(Q^2+S^2)}\frac{\partial}{\partial\phi}],
\end{split}
\end{equation}
whereas $V_g$ is the geometric potential, that is
\begin{equation}\label{GP01}
 V_g=-\frac{\hbar^2}{2m^*}\frac{(Q^2+2S^2-\epsilon N^2Q\cos N\phi)^2}{(Q^2+S^2)^3},
\end{equation}
a function of $R_0$, $\phi$, $\varepsilon$ and $N$. The specific properties are described in Fig.~\ref{corrugated GPv}, the geometric potential wells are dramatically deepened by increasing the magnitude of corrugations, or by increasing the periodic number $N$, even or by decreasing the radius $R_0$. As an important feature of the geometric potential, its minimum is the following form

\begin{equation}\label{minimun gp}
V_{gm}=-\frac{\hbar^2}{2m^*}(\frac{1}{2 R_{0}}-\frac{\epsilon N^2}{(2R_{0})^2})^2.
\end{equation}
The geometric potential versus $\phi$ for $N=even$ and $N=odd$ are described in Figs.~\ref{corrugated GPv} (c) and (d), respectively. It is straightforward to find that the extreme points are located at $\phi=\frac{\pi}{N}\ast i$  $(i=0,1,2,\ldots,N)$. Notice that at $\phi=\pi$ the geometric potential takes the minimum value for $N=even$, the subminimum value for $N=odd$.

It is well known that the geometric potential is induced by the curvature, in the present system the curvature is considerably affected by the corrugations. Specifically, in what follows, we will investigate the electronic energy level structure, corresponding eigenstates and energy shifts affected by the corrugations.

\section{Bound states and energy level structure of confined elections induced by corrugations}
It is distinct that the effective Hamiltonian~\eqref{h1} can be analytically separated into a $z$-component and another describing the dynamics of electron moving along the annular corrugated wire. In the present paper, we mainly focus on the investigation of the effective dynamics of annular wire, do not further discuss that of $z$-axis. If the length in the $z$-direction is much larger than the size of the annular corrugated wire, the $z$-component of energy would be continuous. In the present paper, the considered annular corrugated surface has in general a certain width in the $z$ direction, the $z$-component energy is then quantized. In fact, whether the $z$-component of energy is quantized or not, the remaining dynamics of the annular corrugated wire is considerably affected by the geometric potential induced by the corrugations. In terms of Eqs.~\eqref{MyNote}, ~\eqref{kinetic} and ~\eqref{GP01}, the wave function $\psi(\xi)$ can be replaced by $\psi(\phi)$ with certain $R_0$, $\phi$, $\epsilon$ and $N$.

As shown in Fig.~\ref{corrugated GPv}, it is obvious that the geometric potential is spatial reflectional invariance and periodical invariance with respect to $\phi$. Therefore, the associated eigenfunction also owns the corresponding properties. For the sake of convenience, we take the eigenfunction in the following form
\begin{equation}\label{one dimension wavefunction}
\psi(\phi)=\sum_{k=0}^{\infty}c_{k}\cos(kN\phi).
\end{equation}
According to the wave function~\eqref{one dimension wavefunction}, we diagonalize the effective Hamiltonian $H_{\rm{eff}}$ to obtain the eigenenergies and eigenstates for certain $R_{0}$, $\epsilon$ and $N$. In our calculations, the results are derived by truncating $k$ with $k=100$ that is sufficient for convergence.
\begin{figure}[htbp]\label{energy1}
\includegraphics[width=0.50\textwidth]{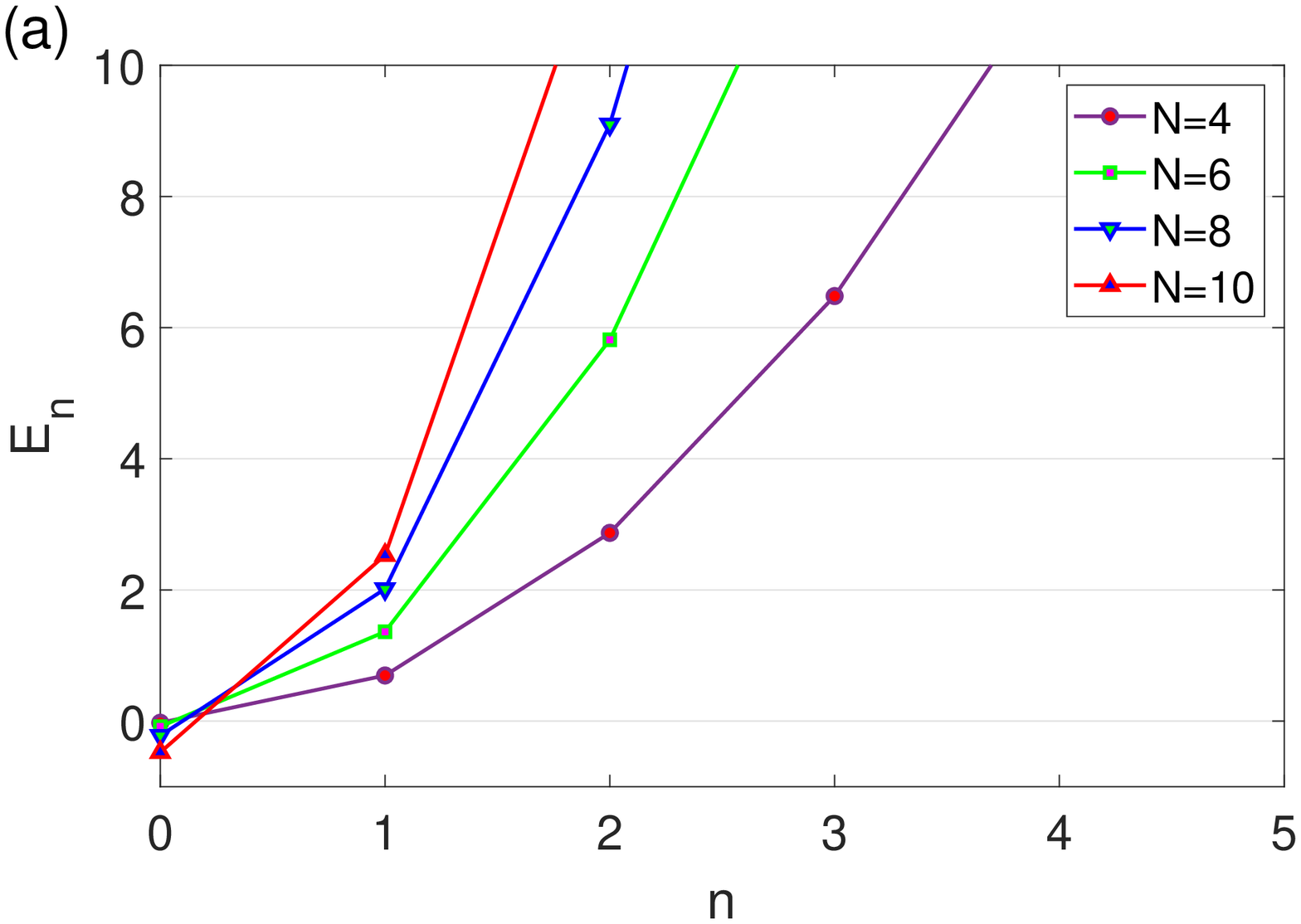}
\includegraphics[width=0.50\textwidth]{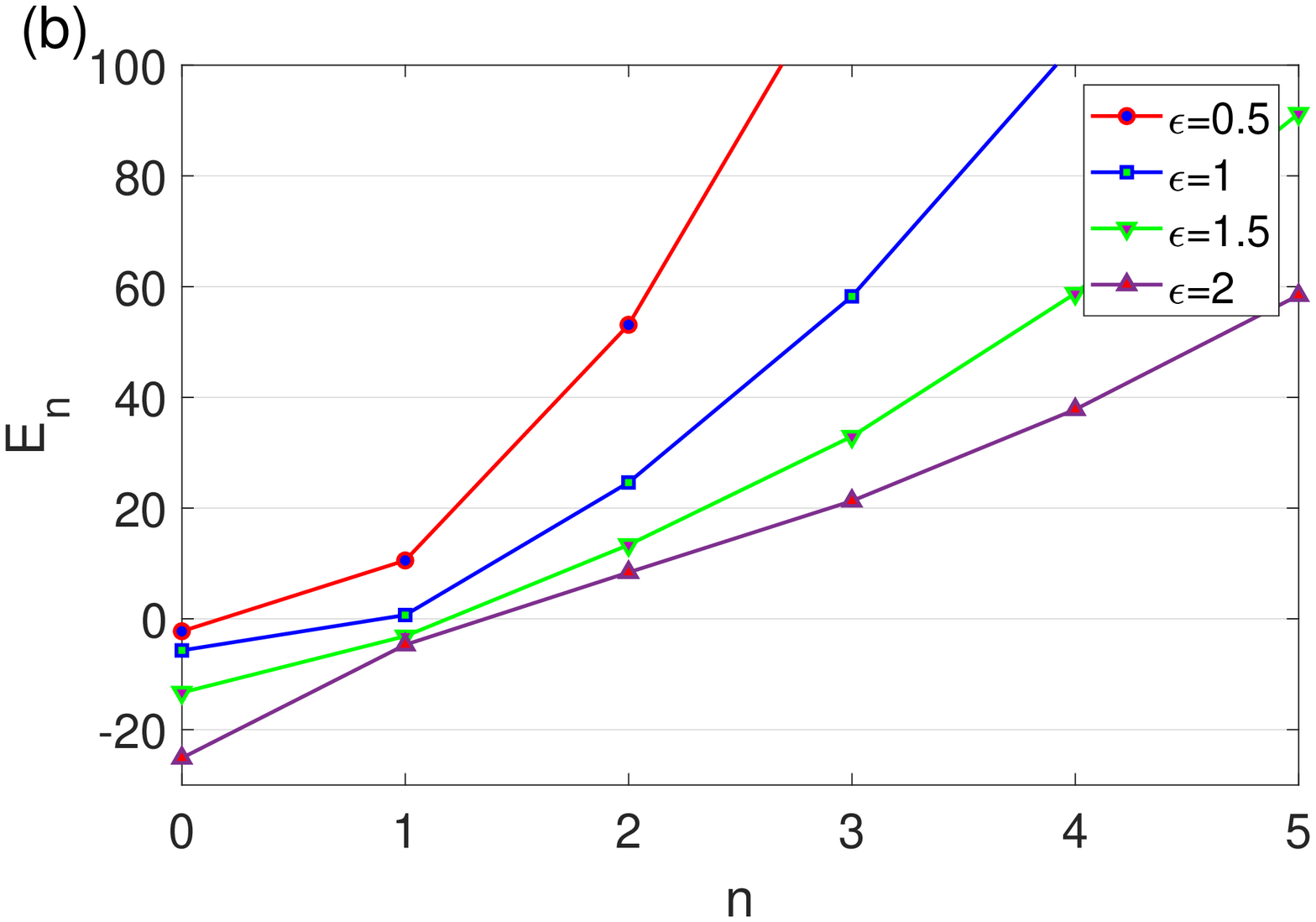}
\caption{\footnotesize (Color online) Electronic energies of an annular corrugated wire with (a) $R_{0}=4$, $\epsilon=1$ and $N=4, 6, 8, 10$, (b) $R_{0}=4$, $N=20$ and $\epsilon=0.5, 1, 1.5, 2$. The points are the numerical results of the exact diagonalization of the $H_{\rm{eff}}$. Here $\frac{\hbar^2}{2m^*l^{2}}$ is taken as an unit.}\label{Fig3}
\end{figure}

\begin{figure}[htbp]\label{energy}
\includegraphics[width=0.50\textwidth]{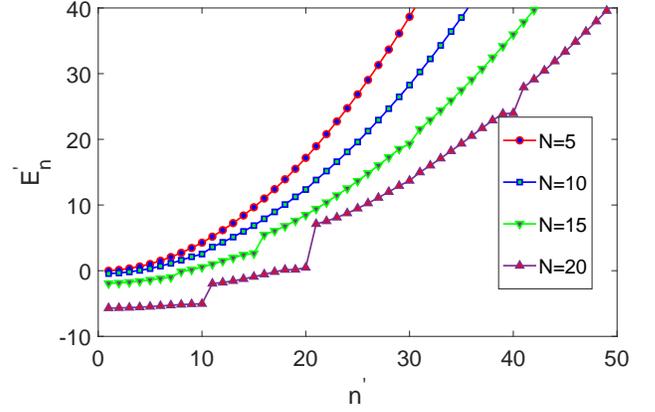}
\caption{\footnotesize (Color online)Electronic energies of an annular corrugated wire based on the eigenfunction of the ring line with $R_{0}=4$, $\epsilon=1$ and $N$=5, 10, 15, 20. The points are the numerical energy levels. Here $\frac{\hbar^2}{2m^*l^{2}}$ is taken as an unit.}\label{Fig4}
\end{figure}

By calculating numerically, the electronic energy levels of an annular corrugated wire are described in Fig.~\ref{Fig3} (a) and (b) for the case of $R_0=4$, $\epsilon=1$, $N=4, 6, 8, 10$ and that of $R_0=4$, $N=20$, $\epsilon=0.5, 1, 1.5, 2$. $n$ and $E_{n}$, label the eigenmodes and the corresponding eigenenergies of the Hamiltonian, where chooses $l$ as a length unit, respectively. As described in Fig.~\ref{Fig3} (a), the transition energies between the adjacent energy levels increase dramatically as the number of corrugations increases. Moreover, there are negative energy levels that refer to $E<0$ states, which denotes that the motion of electron along the annular corrugated wire are constrained by the geometric potential induced by the corrugations. However, the number of negative energy levels does not change with the increase of the number of corrugations, as shown in TABLE I. The presences of the negative eigenenergies and that of the bound states are eventually determined by the attractive wells of the geometric potential $V_g$. The potential wells are sharply deepened and the transition energies are dramatically enlarged by adding corrugations. It is worthwhile to notice that the number of bound states is not added by adding corrugations. Obviously, the potential wells can also be deepened by increasing the magnitude of corrugations. As shown in Fig.~\ref{Fig3}(b), the number of negative eigenenergies and that of bound states are added by enlarging the magnitude of corrugations. As a consequence, the negative eigenenergies and bound states can be produced by designing corrugations for an annular wire. Practically, these results are helpful to control the electronic transport by designing particular geometries~\cite{Moraes2016Geometric,Wang2016Transmission,condmat4010003,Cao2019}.

\begin{table}
 \tb{TABLE I.The number of negative energy level \tb{$N_{E}$} and bound states \tb{$N_{B}$}, the eigenvalues of the ground states \tb{$E_{0}$} and the transition energy between the ground state and first excited state \tb{$\Delta E_{1}$} for the annular wire of $R_0=4$, $\epsilon=1$ and $N=4, 6, 8, 10$.}
  \renewcommand\tabcolsep{7.0pt}
\begin{tabular}{ccccccc}
  \hline
  {N} & {$\epsilon$} & {R} &{$N_{E}$} &{$N_{B}$} &{$E_{0}$} &{$\Delta E_{1}$}  \\
    4 & 1 & 4 &1 &1 & -0.02587295 & 0.72118644 \\
    6 & 1 & 4 &1 &1 & -0.08365887 & 1.44595252 \\
    8 & 1 & 4 &1 &1 & -0.22147269 & 2.23720045 \\
    10& 1 & 4 &1 &1 & -0.47715353 & 3.00619753 \\
  \hline
\end{tabular}
\end{table}

\begin{table}
 \tb{TABLE II. The first excited state energy \tb{$ E_{1}$} and the second excited state energy \tb{$ E_{2}$} of annular corrugated wire of $R_0=4$, $\epsilon=1$ and $N=5, 10, 15, 20$ and the energy level \tb{$ E^{'}_{N}$} and \tb{$ E^{'}_{2N}$} based on the wavefunction of annular wire.}
\begin{tabular}{ccccccc}
  \hline
   {N} & {$\varepsilon$} & {R} &{$ E_{1}$} &{$ E^{'}_{N}$} &{$E_{2}$} &{$ E^{'}_{2N}$}  \\
    5  & 1 & 4   & 1.02076686 & 1.02006811 & 4.27138058  & 4.27122914 \\
    10 & 1 & 4   & 2.52904400 & 2.52833661 & 12.37233876 & 12.35684566 \\
    15 & 1 & 4   & 2.68997342 & 2.63327893 & 19.46819019 & 19.31057811 \\
    20 & 1 & 4   & 0.68449421 & 0.47127916 & 24.62154144 & 23.98627050 \\
  \hline
\end{tabular}
\end{table}

Another important result in the present paper is that the relationship between the transition energy and the number of corrugations. In order to investigate the relation, we numerically calculate the electronic energy levels for an annular corrugated wire of $R_0=4$, $\epsilon=1$ with $N=5, 10, 15, 20$ by the wavefunctions Eq.~\eqref{one dimension wavefunction} presented in Fig.~\ref{Fig3} and $cos(k\phi)$, eigenfunction of annular wire without corrugations, as shown in Fig.~\ref{Fig4} and TABLE II. $n^{'}$ is the modes and $E^{'}_{n}$ is the energy levels. It is easy to find that the significant energy level transition usually happens between $E^{'}_N$ and $E^{'}_{N+1}$ as shown in Fig.~\ref{Fig4}, which denotes that the eigenenergies of the annular corrugated wire are located at $n^{'}=m\ast N (m=1,2,\cdots)$, as shown in TABLE II. Therefore, the transition energy between the adjacent eigenstates of the annular wire with $N$ corrugations corresponds to $N$ energy levels differences. That is to say, for the annular wire with $N$ corrugations, each transition energy between the adjacent eigenstates is capable of accommodating $N$ energy levels of the annular wire without corrugations. As a consequence, the transition energies are considerably increased by adding corrugations.

\section{Geometric effects on energy shifts and the ground state probability density distribution}
For the further investigation of the geometric effects on the effective dynamics, we will calculate the geometric energy shift of the ground state and the ground state probability density distribution numerically. For simplicity, the energy shift is defined as
\begin{equation}\label{energy shift}
\Delta E_{V_{g}}=\langle \psi_{0}|V_{g}|\psi_{0}\rangle
\end{equation}
by the wave function of ground state $|\psi_0\rangle$ and the geometric potential $V_{g}$.

As shown in TABLE III and IV, it is obvious that the geometrically induced energy shift is very important to the effective dynamics, and even the absolute value of the energy shift is greater than that of the ground state energy. It can be confirmed that the energies of the ground states are considerably shifted by the geometric potential. Since the curvature-induced potential wells are dramatically deepened by increasing the number of corrugations $N$ and the magnitude $\epsilon$, as a result, the bound states can be produced by adding the number of corrugations $N$ or enlarging the magnitude $\epsilon$ for the annular corrugated wire. As a consequence, a stronger geometric potential can result in lower bound states, which can make the resonances peaks in the electronic transmittance be shifted to lower energies practically.
\begin{table}
 \tb{TABLE III. The ground state energy \tb{$E_{0}$} and the energy shift \tb{$\Delta E_{V_{g}}$ for the annular wire of $R_0=4$, $\epsilon=1$ and $N=5, 10, 15, 20$}.}
  \renewcommand\tabcolsep{5.0pt}
  \begin{tabular}{ccccc}
    \hline
    {N} & {$\varepsilon$} & {$R_{0}$} & {$E_{0}$[$unit$ $of$ $\frac{\hbar^2}{2m^*l^{2}}$]} &{$\Delta E_{V_{g}}$[$unit$ $of$ $\frac{\hbar^2}{2m^*l^{2}}$]$$} \\
    5 & 1 & 4 & -0.04701866 & -0.0478929590086791 \\
    10 & 1 & 4 & -0.47715353 & -0.495665874076816    \\
    15 & 1 & 4 & -1.95698662 & -2.23824342373883   \\
    20 & 1 & 4 & -5.69520598 & -7.56689752279481   \\
    \hline
  \end{tabular}
  \label{table1}
\end{table}

\begin{table}
 \tb{TABLE IV. The ground state energy \tb{$E_{0}$} and the energy shift \tb{$\Delta E_{V_{g}}$} for the annular wire of $R_0=4$, $N=20$ and $\epsilon=0.5, 1, 1.5, 2$.}
 \renewcommand\tabcolsep{5.0pt}
  \begin{tabular}{ccccc}
    \hline
    {$\varepsilon$} & {$R_{0}$} & {N} & {$E_{0}$[$unit$ $of$ $\frac{\hbar^2}{2m^*l^{2}}$]} &{$\Delta E_{V_{g}}$[$unit$ $of$ $\frac{\hbar^2}{2m^*l^{2}}$]$$}  \\
    0.5 & 4 & 20 & -2.24279515 & -2.33942875912832  \\
    1   & 4 & 20 &  -5.69520598 & -7.56689752279481    \\
    1.5 & 4 & 20 &  -13.31164212 & -20.4801652650634   \\
    2   & 4 & 20 & -25.13234603 & -38.0561428901224 \\
    \hline
  \end{tabular}
  \label{table2}
\end{table}

\begin{figure}[htbp]\label{psi}
\includegraphics[width=0.44\textwidth]{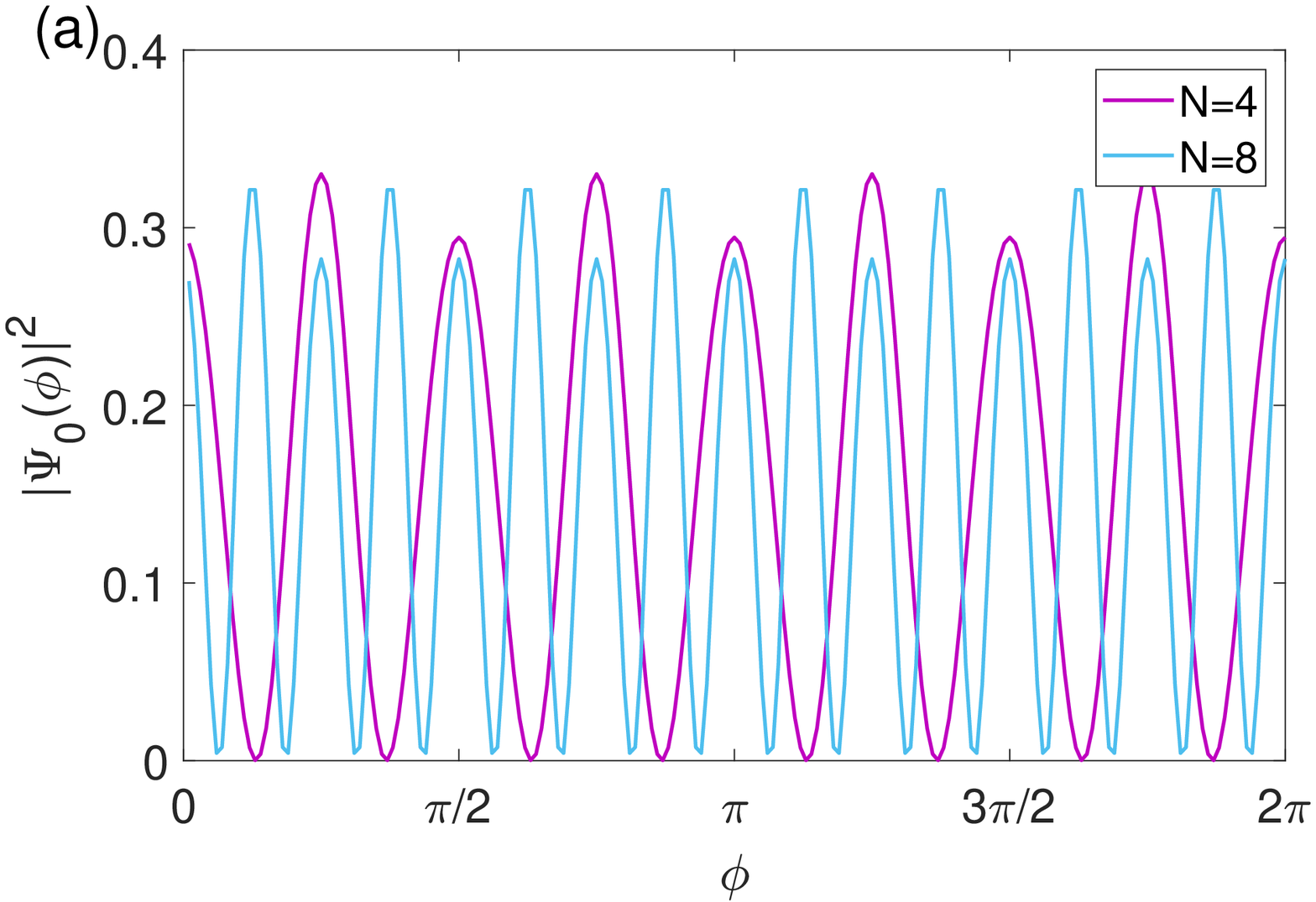}
\includegraphics[width=0.44\textwidth]{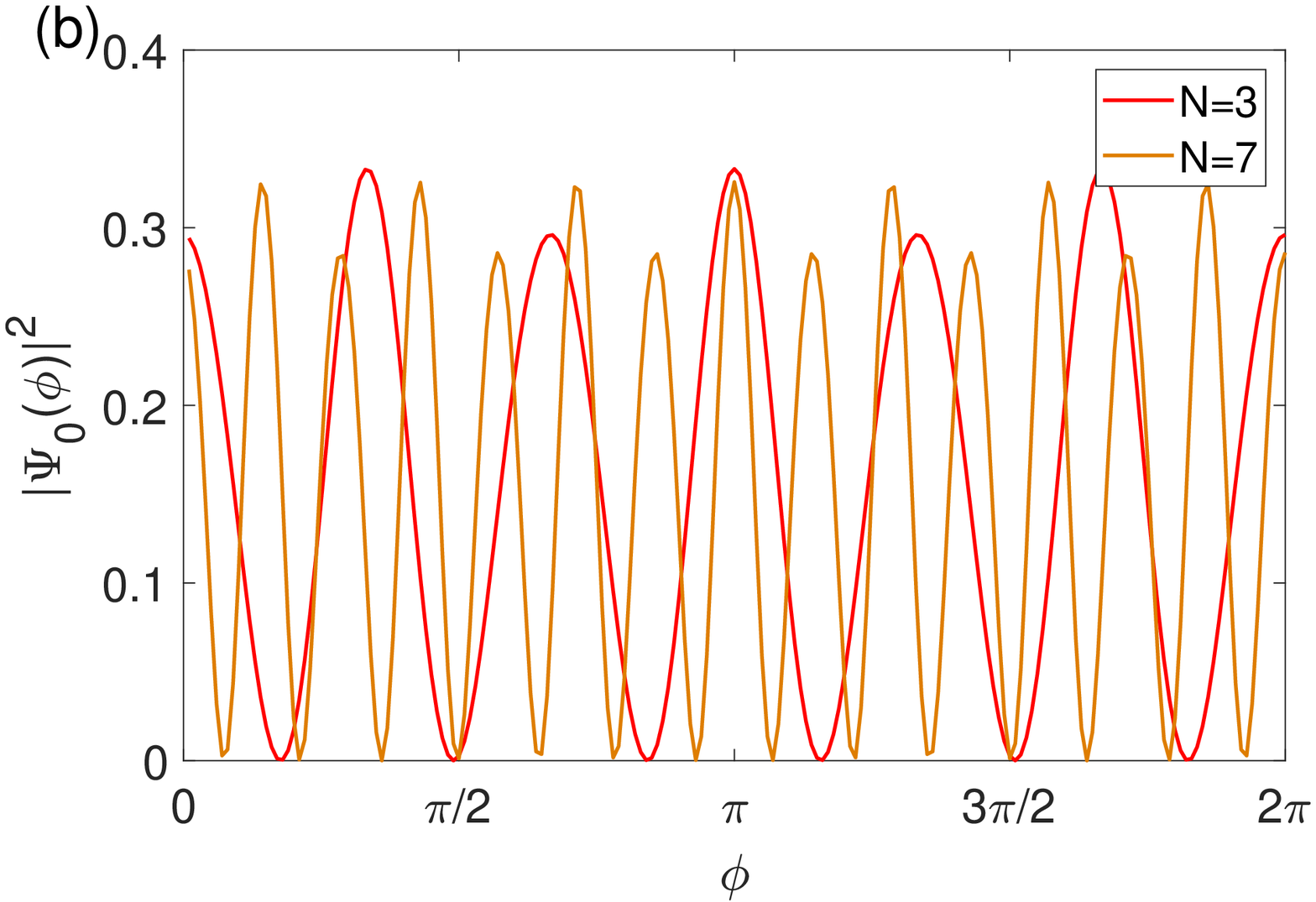}
\includegraphics[width=0.44\textwidth]{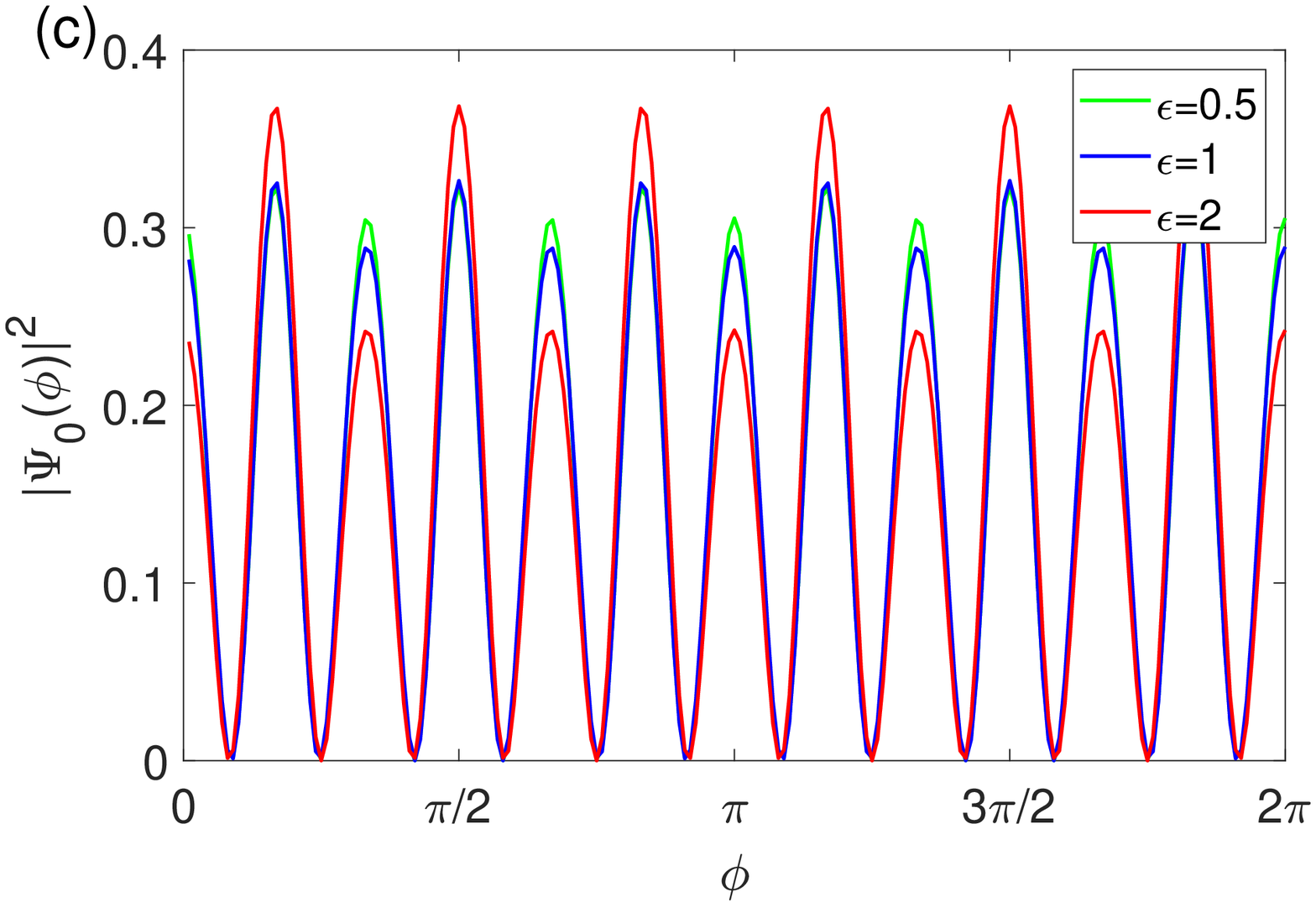}
\caption{\footnotesize (Color online) The ground state probability density as a function of azimuthal angle $\phi$ with (a) $R_{0}=4$, $\epsilon=1$ and $N=4, 8$, (b) $R_{0}=4$, $\epsilon=1$ and $N=3, 7$, (c) $R_{0}=4$, $\epsilon=1$ and $N=3, 4$, (d) $R_{0}=4$, $N=6$ and $\epsilon=0.5, 1, 2$.}\label{Fig5}
\end{figure}

Describing a quantum system that requires not only an effective Hamiltonian but also the wave functions. For simplicity, we just consider the ground state probability density distribution by calculating numerically as described in Fig.~\ref{Fig5}. It is apparent that the ground state probability density is relatively more concentrated as the number of corrugations increases as shown in Figs.~\ref{Fig5} (a) and (b). Combining Fig.~\ref{Fig5} with Fig.~\ref{corrugated GPv}, it is easy to find that the number of probability density peaks is in nice agreement with the number of the geometric potential wells. At $\phi=\frac{\pi}{N}\ast 2i$  $(i=0,1,2,\ldots,N)$, the probability density takes its submaximum values, while the geometric potential is minimum. At $\phi=\frac{\pi}{N}\ast (2i+1)$  $(i=0,1,2,\ldots,N)$, while the probability density is maximum, the geometric potential takes its subminimum values. In addition, at $\phi=\pi$, the probability density gradually decreases as the number of corrugations increases as shown in Figs.~\ref{Fig5} (a) and (b). According to Fig.~\ref{Fig5} (c), with the increase of the magnitude of corrugations, the maximal ground state probability density gradually decreases, while the results of the submaximal probability demsity are opposite.  As a conclusion, the geometric effects not only affect the equivalent mechanical quantity operator, but also reconstruct the probability density distribution of the ground state.

\section{CONCLUSION AND DISCUSSION}
In this paper, we have considered a quantum dynamical system describing electrons without interactions confined to an annular corrugated surface. We directly gave the effective Hamiltonian by using the thin-layer quantization approach in which the geometric potential plays an important role. The geometric potential induced by curvature consists of attractive wells with different depthes that are reconstructed by corrugations. Specifically, the attractive wells are dramatically deepened as the number of corrugations increases or the magnitude does. In particular, the number of deepest wells nicely agree with the number of corrugations, and the number of second deep wells is also the same. What's more, the geometric potential is still valid to a partially deformed annular wire with $\phi$ ranging from $\phi_{0}$ to $\phi_{1}$ $(0\leq\phi_{0}<\phi_{1}\leq2\pi)$.

In terms of the effective dynamics, we have investigated the electronic energy level structure and the corresponding eigenstates, found that the number of bound states can be added by enlarging the magnitude of corrugations, and the transition energies can be increased by adding the number of corrugations. In particular, the transition energy between the adjacent eigenstates of the annular corrugated wire, corresponds to $N$ energy levels difference based on the wavefunction of annular wire, and the number of the energy levels is equal to the number of corrugations. In the investigation procedure, the geometric potential plays an important role in the effective dynamics and the bound states. The energy shift of the ground state contributed by the geometric potential owns an absolute value larger than that of the ground state energy. And the ground state probability density distribution is also redistributed by the geometric potential.

As application potentials, the corrugations can be employed to improve the connectivity of nanotubes, to provide new bound states, to reconstruct the electronic energy level structure, and to redistribute the bound states density. These results provide a way to design nanotube-based electronic and photonic devices by introducing corrugations.

\section*{Acknowledgments}
This work is jointly supported by the National Major state Basic Research and Development of China (Grant No. 2016YFE0129300), the National Nature Science Foundation of China (Grants No. 11690030, No. 11475085, No. 11535005, No. 61425018). Y.-L. W. was funded by the Natural Science Foundation of Shandong Province of China (Grant No. ZR2017MA010).

\normalem
\bibliographystyle{apsrev4-1}
\bibliography{chengrun}

\begin{thebibliography}{36}%
\makeatletter
\providecommand \@ifxundefined [1]{%
 \@ifx{#1\undefined}
}%
\providecommand \@ifnum [1]{%
 \ifnum #1\expandafter \@firstoftwo
 \else \expandafter \@secondoftwo
 \fi
}%
\providecommand \@ifx [1]{%
 \ifx #1\expandafter \@firstoftwo
 \else \expandafter \@secondoftwo
 \fi
}%
\providecommand \natexlab [1]{#1}%
\providecommand \enquote  [1]{``#1''}%
\providecommand \bibnamefont  [1]{#1}%
\providecommand \bibfnamefont [1]{#1}%
\providecommand \citenamefont [1]{#1}%
\providecommand \href@noop [0]{\@secondoftwo}%
\providecommand \href [0]{\begingroup \@sanitize@url \@href}%
\providecommand \@href[1]{\@@startlink{#1}\@@href}%
\providecommand \@@href[1]{\endgroup#1\@@endlink}%
\providecommand \@sanitize@url [0]{\catcode `\\12\catcode `\$12\catcode
  `\&12\catcode `\#12\catcode `\^12\catcode `\_12\catcode `\%12\relax}%
\providecommand \@@startlink[1]{}%
\providecommand \@@endlink[0]{}%
\providecommand \url  [0]{\begingroup\@sanitize@url \@url }%
\providecommand \@url [1]{\endgroup\@href {#1}{\urlprefix }}%
\providecommand \urlprefix  [0]{URL }%
\providecommand \Eprint [0]{\href }%
\providecommand \doibase [0]{http://dx.doi.org/}%
\providecommand \selectlanguage [0]{\@gobble}%
\providecommand \bibinfo  [0]{\@secondoftwo}%
\providecommand \bibfield  [0]{\@secondoftwo}%
\providecommand \translation [1]{[#1]}%
\providecommand \BibitemOpen [0]{}%
\providecommand \bibitemStop [0]{}%
\providecommand \bibitemNoStop [0]{.\EOS\space}%
\providecommand \EOS [0]{\spacefactor3000\relax}%
\providecommand \BibitemShut  [1]{\csname bibitem#1\endcsname}%
\let\auto@bib@innerbib\@empty
\bibitem [{\citenamefont {Goldstone}\ and\ \citenamefont
  {Jaffe}(1992)}]{Goldstone1992}%
  \BibitemOpen
  \bibfield  {author} {\bibinfo {author} {\bibfnamefont {J.}~\bibnamefont
  {Goldstone}}\ and\ \bibinfo {author} {\bibfnamefont {R.~L.}\ \bibnamefont
  {Jaffe}},\ }\href {\doibase 10.1103/PhysRevB.45.14100} {\bibfield  {journal}
  {\bibinfo  {journal} {Phys. Rev. B}\ }\textbf {\bibinfo {volume} {45}},\
  \bibinfo {pages} {14100} (\bibinfo {year} {1992})}\BibitemShut {NoStop}%
\bibitem [{\citenamefont {Ouyang}\ \emph {et~al.}(1999)\citenamefont {Ouyang},
  \citenamefont {Mohta},\ and\ \citenamefont {Jaffe}}]{Ouyang1999}%
  \BibitemOpen
  \bibfield  {author} {\bibinfo {author} {\bibfnamefont {P.}~\bibnamefont
  {Ouyang}}, \bibinfo {author} {\bibfnamefont {V.}~\bibnamefont {Mohta}}, \
  and\ \bibinfo {author} {\bibfnamefont {R.}~\bibnamefont {Jaffe}},\ }\href
  {\doibase http://dx.doi.org/10.1006/aphy.1999.5935} {\bibfield  {journal}
  {\bibinfo  {journal} {Ann. Phys.}\ }\textbf {\bibinfo {volume} {275}},\
  \bibinfo {pages} {297 } (\bibinfo {year} {1999})}\BibitemShut {NoStop}%
\bibitem [{\citenamefont {Shima}\ \emph {et~al.}(2010)\citenamefont {Shima},
  \citenamefont {Sato}, \citenamefont {Iiboshi}, \citenamefont {Ghosh},\ and\
  \citenamefont {Arroyo}}]{Shima2010}%
  \BibitemOpen
  \bibfield  {author} {\bibinfo {author} {\bibfnamefont {H.}~\bibnamefont
  {Shima}}, \bibinfo {author} {\bibfnamefont {M.}~\bibnamefont {Sato}},
  \bibinfo {author} {\bibfnamefont {K.}~\bibnamefont {Iiboshi}}, \bibinfo
  {author} {\bibfnamefont {S.}~\bibnamefont {Ghosh}}, \ and\ \bibinfo {author}
  {\bibfnamefont {M.}~\bibnamefont {Arroyo}},\ }\href {\doibase
  10.1103/PhysRevB.82.085401} {\bibfield  {journal} {\bibinfo  {journal} {Phys.
  Rev. B}\ }\textbf {\bibinfo {volume} {82}},\ \bibinfo {pages} {085401}
  (\bibinfo {year} {2010})}\BibitemShut {NoStop}%
\bibitem [{\citenamefont {Novakovic}\ \emph {et~al.}(2011)\citenamefont
  {Novakovic}, \citenamefont {Akis},\ and\ \citenamefont
  {Knezevic}}]{Novakovic2011}%
  \BibitemOpen
  \bibfield  {author} {\bibinfo {author} {\bibfnamefont {B.}~\bibnamefont
  {Novakovic}}, \bibinfo {author} {\bibfnamefont {R.}~\bibnamefont {Akis}}, \
  and\ \bibinfo {author} {\bibfnamefont {I.}~\bibnamefont {Knezevic}},\ }\href
  {\doibase 10.1103/PhysRevB.84.195419} {\bibfield  {journal} {\bibinfo
  {journal} {Phys. Rev. B}\ }\textbf {\bibinfo {volume} {84}},\ \bibinfo
  {pages} {195419} (\bibinfo {year} {2011})}\BibitemShut {NoStop}%
\bibitem [{\citenamefont {Santos}\ \emph {et~al.}(2016)\citenamefont {Santos},
  \citenamefont {Fumeron}, \citenamefont {Berche},\ and\ \citenamefont
  {Moraes}}]{Moraes2016Geometric}%
  \BibitemOpen
  \bibfield  {author} {\bibinfo {author} {\bibfnamefont {F.}~\bibnamefont
  {Santos}}, \bibinfo {author} {\bibfnamefont {S.}~\bibnamefont {Fumeron}},
  \bibinfo {author} {\bibfnamefont {B.}~\bibnamefont {Berche}}, \ and\ \bibinfo
  {author} {\bibfnamefont {F.}~\bibnamefont {Moraes}},\ }\href
  {http://stacks.iop.org/0957-4484/27/i=13/a=135302} {\bibfield  {journal}
  {\bibinfo  {journal} {Nanotechnology}\ }\textbf {\bibinfo {volume} {27}},\
  \bibinfo {pages} {135302} (\bibinfo {year} {2016})}\BibitemShut {NoStop}%
\bibitem [{\citenamefont {Cheng}\ \emph {et~al.}(2019)\citenamefont {Cheng},
  \citenamefont {Wang}, \citenamefont {Jiang}, \citenamefont {Liu},\ and\
  \citenamefont {Zong}}]{condmat4010003}%
  \BibitemOpen
  \bibfield  {author} {\bibinfo {author} {\bibfnamefont {R.}~\bibnamefont
  {Cheng}}, \bibinfo {author} {\bibfnamefont {Y.-L.}\ \bibnamefont {Wang}},
  \bibinfo {author} {\bibfnamefont {H.}~\bibnamefont {Jiang}}, \bibinfo
  {author} {\bibfnamefont {X.-J.}\ \bibnamefont {Liu}}, \ and\ \bibinfo
  {author} {\bibfnamefont {H.-S.}\ \bibnamefont {Zong}},\ }\href
  {http://www.mdpi.com/2410-3896/4/1/3} {\bibfield  {journal} {\bibinfo
  {journal} {Condens. Matter}\ }\textbf {\bibinfo {volume} {4}} (\bibinfo
  {year} {2019})}\BibitemShut {NoStop}%
\bibitem [{\citenamefont {Ortix}\ and\ \citenamefont {van~den
  Brink}(2010{\natexlab{a}})}]{Ortix2010}%
  \BibitemOpen
  \bibfield  {author} {\bibinfo {author} {\bibfnamefont {C.}~\bibnamefont
  {Ortix}}\ and\ \bibinfo {author} {\bibfnamefont {J.}~\bibnamefont {van~den
  Brink}},\ }\href {\doibase 10.1103/PhysRevB.81.165419} {\bibfield  {journal}
  {\bibinfo  {journal} {Phys. Rev. B}\ }\textbf {\bibinfo {volume} {81}},\
  \bibinfo {pages} {165419} (\bibinfo {year} {2010}{\natexlab{a}})}\BibitemShut
  {NoStop}%
\bibitem [{\citenamefont {Tanda}\ \emph {et~al.}(2002)\citenamefont {Tanda},
  \citenamefont {Tsuneta}, \citenamefont {Okajima}, \citenamefont {Inagaki},
  \citenamefont {Yamaya},\ and\ \citenamefont {Hatakenaka}}]{Tanda2002}%
  \BibitemOpen
  \bibfield  {author} {\bibinfo {author} {\bibfnamefont {S.}~\bibnamefont
  {Tanda}}, \bibinfo {author} {\bibfnamefont {T.}~\bibnamefont {Tsuneta}},
  \bibinfo {author} {\bibfnamefont {Y.}~\bibnamefont {Okajima}}, \bibinfo
  {author} {\bibfnamefont {K.}~\bibnamefont {Inagaki}}, \bibinfo {author}
  {\bibfnamefont {K.}~\bibnamefont {Yamaya}}, \ and\ \bibinfo {author}
  {\bibfnamefont {N.}~\bibnamefont {Hatakenaka}},\ }\href
  {https://doi.org/10.1038/417397a} {\bibfield  {journal} {\bibinfo  {journal}
  {Nature}\ }\textbf {\bibinfo {volume} {417}},\ \bibinfo {pages} {397}
  (\bibinfo {year} {2002})}\BibitemShut {NoStop}%
\bibitem [{\citenamefont {Gravesen}\ \emph {et~al.}(2005)\citenamefont
  {Gravesen}, \citenamefont {Willatzen},\ and\ \citenamefont
  {Voon}}]{Gravesen2005}%
  \BibitemOpen
  \bibfield  {author} {\bibinfo {author} {\bibfnamefont {J.}~\bibnamefont
  {Gravesen}}, \bibinfo {author} {\bibfnamefont {M.}~\bibnamefont {Willatzen}},
  \ and\ \bibinfo {author} {\bibfnamefont {L.~C. L.~Y.}\ \bibnamefont {Voon}},\
  }\href {http://dx.doi.org/10.1063/1.1829376} {\bibfield  {journal} {\bibinfo
  {journal} {J. Math. Phys.}\ }\textbf {\bibinfo {volume} {46}},\ \bibinfo
  {pages} {012107} (\bibinfo {year} {2005})}\BibitemShut {NoStop}%
\bibitem [{\citenamefont {da~Costa}(1981)}]{Costa1981}%
  \BibitemOpen
  \bibfield  {author} {\bibinfo {author} {\bibfnamefont {R.~C.~T.}\
  \bibnamefont {da~Costa}},\ }\href {\doibase 10.1103/PhysRevA.23.1982}
  {\bibfield  {journal} {\bibinfo  {journal} {Phys. Rev. A}\ }\textbf {\bibinfo
  {volume} {23}},\ \bibinfo {pages} {1982} (\bibinfo {year}
  {1981})}\BibitemShut {NoStop}%
\bibitem [{\citenamefont {Schuster}\ and\ \citenamefont
  {Jaffe}(2003)}]{Jaffe2003Quantum}%
  \BibitemOpen
  \bibfield  {author} {\bibinfo {author} {\bibfnamefont {P.}~\bibnamefont
  {Schuster}}\ and\ \bibinfo {author} {\bibfnamefont {R.}~\bibnamefont
  {Jaffe}},\ }\href {\doibase http://dx.doi.org/10.1016/S0003-4916(03)00080-0}
  {\bibfield  {journal} {\bibinfo  {journal} {Ann. Phys.}\ }\textbf {\bibinfo
  {volume} {307}},\ \bibinfo {pages} {132 } (\bibinfo {year}
  {2003})}\BibitemShut {NoStop}%
\bibitem [{\citenamefont {Cantele}\ \emph {et~al.}(2000)\citenamefont
  {Cantele}, \citenamefont {Ninno},\ and\ \citenamefont
  {Iadonisi}}]{Cantele2000Topological}%
  \BibitemOpen
  \bibfield  {author} {\bibinfo {author} {\bibfnamefont {G.}~\bibnamefont
  {Cantele}}, \bibinfo {author} {\bibfnamefont {D.}~\bibnamefont {Ninno}}, \
  and\ \bibinfo {author} {\bibfnamefont {G.}~\bibnamefont {Iadonisi}},\ }\href
  {\doibase 10.1103/PhysRevB.61.13730} {\bibfield  {journal} {\bibinfo
  {journal} {Phys. Rev. B}\ }\textbf {\bibinfo {volume} {61}},\ \bibinfo
  {pages} {13730} (\bibinfo {year} {2000})}\BibitemShut {NoStop}%
\bibitem [{\citenamefont {Aoki}\ \emph {et~al.}(2001)\citenamefont {Aoki},
  \citenamefont {Koshino}, \citenamefont {Takeda}, \citenamefont {Morise},\
  and\ \citenamefont {Kuroki}}]{Aoki2001Electronic}%
  \BibitemOpen
  \bibfield  {author} {\bibinfo {author} {\bibfnamefont {H.}~\bibnamefont
  {Aoki}}, \bibinfo {author} {\bibfnamefont {M.}~\bibnamefont {Koshino}},
  \bibinfo {author} {\bibfnamefont {D.}~\bibnamefont {Takeda}}, \bibinfo
  {author} {\bibfnamefont {H.}~\bibnamefont {Morise}}, \ and\ \bibinfo {author}
  {\bibfnamefont {K.}~\bibnamefont {Kuroki}},\ }\href {\doibase
  10.1103/PhysRevB.65.035102} {\bibfield  {journal} {\bibinfo  {journal} {Phys.
  Rev. B}\ }\textbf {\bibinfo {volume} {65}},\ \bibinfo {pages} {035102}
  (\bibinfo {year} {2001})}\BibitemShut {NoStop}%
\bibitem [{\citenamefont {Encinosa}\ and\ \citenamefont
  {Mott}(2003)}]{Encinosa2003}%
  \BibitemOpen
  \bibfield  {author} {\bibinfo {author} {\bibfnamefont {M.}~\bibnamefont
  {Encinosa}}\ and\ \bibinfo {author} {\bibfnamefont {L.}~\bibnamefont
  {Mott}},\ }\href {\doibase 10.1103/PhysRevA.68.014102} {\bibfield  {journal}
  {\bibinfo  {journal} {Phys. Rev. A}\ }\textbf {\bibinfo {volume} {68}},\
  \bibinfo {pages} {014102} (\bibinfo {year} {2003})}\BibitemShut {NoStop}%
\bibitem [{\citenamefont {Taira}\ and\ \citenamefont
  {Shima}(2007{\natexlab{a}})}]{Taira2007Curvature}%
  \BibitemOpen
  \bibfield  {author} {\bibinfo {author} {\bibfnamefont {H.}~\bibnamefont
  {Taira}}\ and\ \bibinfo {author} {\bibfnamefont {H.}~\bibnamefont {Shima}},\
  }\href {\doibase https://doi.org/10.1016/j.susc.2007.04.220} {\bibfield
  {journal} {\bibinfo  {journal} {Surf. Sci.}\ }\textbf {\bibinfo {volume}
  {601}},\ \bibinfo {pages} {5270 } (\bibinfo {year}
  {2007}{\natexlab{a}})}\BibitemShut {NoStop}%
\bibitem [{\citenamefont {Taira}\ and\ \citenamefont
  {Shima}(2007{\natexlab{b}})}]{Taira2007Electronic}%
  \BibitemOpen
  \bibfield  {author} {\bibinfo {author} {\bibfnamefont {H.}~\bibnamefont
  {Taira}}\ and\ \bibinfo {author} {\bibfnamefont {H.}~\bibnamefont {Shima}},\
  }\href {http://stacks.iop.org/1742-6596/61/i=1/a=226} {\bibfield  {journal}
  {\bibinfo  {journal} {J. Phys.: Conf. Ser.}\ }\textbf {\bibinfo {volume}
  {61}},\ \bibinfo {pages} {1142} (\bibinfo {year}
  {2007}{\natexlab{b}})}\BibitemShut {NoStop}%
\bibitem [{\citenamefont {Ortix}\ and\ \citenamefont {van~den
  Brink}(2010{\natexlab{b}})}]{Ortix2010Effect}%
  \BibitemOpen
  \bibfield  {author} {\bibinfo {author} {\bibfnamefont {C.}~\bibnamefont
  {Ortix}}\ and\ \bibinfo {author} {\bibfnamefont {J.}~\bibnamefont {van~den
  Brink}},\ }\href {\doibase 10.1103/PhysRevB.81.165419} {\bibfield  {journal}
  {\bibinfo  {journal} {Phys. Rev. B}\ }\textbf {\bibinfo {volume} {81}},\
  \bibinfo {pages} {165419} (\bibinfo {year} {2010}{\natexlab{b}})}\BibitemShut
  {NoStop}%
\bibitem [{\citenamefont {Du}\ \emph {et~al.}(2016)\citenamefont {Du},
  \citenamefont {Wang}, \citenamefont {Liang}, \citenamefont {Kang},
  \citenamefont {Liu},\ and\ \citenamefont {Zong}}]{DU201628}%
  \BibitemOpen
  \bibfield  {author} {\bibinfo {author} {\bibfnamefont {L.}~\bibnamefont
  {Du}}, \bibinfo {author} {\bibfnamefont {Y.-L.}\ \bibnamefont {Wang}},
  \bibinfo {author} {\bibfnamefont {G.-H.}\ \bibnamefont {Liang}}, \bibinfo
  {author} {\bibfnamefont {G.-Z.}\ \bibnamefont {Kang}}, \bibinfo {author}
  {\bibfnamefont {X.-J.}\ \bibnamefont {Liu}}, \ and\ \bibinfo {author}
  {\bibfnamefont {H.-S.}\ \bibnamefont {Zong}},\ }\href {\doibase
  https://doi.org/10.1016/j.physe.2015.10.011} {\bibfield  {journal} {\bibinfo
  {journal} {Physica E}\ }\textbf {\bibinfo {volume} {76}},\ \bibinfo {pages}
  {28 } (\bibinfo {year} {2016})}\BibitemShut {NoStop}%
\bibitem [{\citenamefont {Exner}\ \emph {et~al.}(2019)\citenamefont {Exner},
  \citenamefont {Šeba}, \citenamefont {Tater},\ and\ \citenamefont
  {Vaněk}}]{Exner2019a}%
  \BibitemOpen
  \bibfield  {author} {\bibinfo {author} {\bibfnamefont {P.}~\bibnamefont
  {Exner}}, \bibinfo {author} {\bibfnamefont {P.}~\bibnamefont {Šeba}},
  \bibinfo {author} {\bibfnamefont {M.}~\bibnamefont {Tater}}, \ and\ \bibinfo
  {author} {\bibfnamefont {D.}~\bibnamefont {Vaněk}},\ }\href {\doibase
  10.1063/1.531673} {\bibfield  {journal} {\bibinfo  {journal} {J. Math.
  Phys.}\ }\textbf {\bibinfo {volume} {37}},\ \bibinfo {pages} {4867} (\bibinfo
  {year} {2019})}\BibitemShut {NoStop}%
\bibitem [{\citenamefont {Fujita}\ and\ \citenamefont
  {Terasaki}(2005)}]{Fujita2005Band}%
  \BibitemOpen
  \bibfield  {author} {\bibinfo {author} {\bibfnamefont {N.}~\bibnamefont
  {Fujita}}\ and\ \bibinfo {author} {\bibfnamefont {O.}~\bibnamefont
  {Terasaki}},\ }\href {\doibase 10.1103/PhysRevB.72.085459} {\bibfield
  {journal} {\bibinfo  {journal} {Phys. Rev. B}\ }\textbf {\bibinfo {volume}
  {72}},\ \bibinfo {pages} {085459} (\bibinfo {year} {2005})}\BibitemShut
  {NoStop}%
\bibitem [{\citenamefont {Koshino}\ and\ \citenamefont
  {Aoki}(2005)}]{Aoki2005Electronic}%
  \BibitemOpen
  \bibfield  {author} {\bibinfo {author} {\bibfnamefont {M.}~\bibnamefont
  {Koshino}}\ and\ \bibinfo {author} {\bibfnamefont {H.}~\bibnamefont {Aoki}},\
  }\href {\doibase 10.1103/PhysRevB.71.073405} {\bibfield  {journal} {\bibinfo
  {journal} {Phys. Rev. B}\ }\textbf {\bibinfo {volume} {71}},\ \bibinfo
  {pages} {073405} (\bibinfo {year} {2005})}\BibitemShut {NoStop}%
\bibitem [{\citenamefont {Chen}\ and\ \citenamefont
  {Tao}(2009)}]{Chen2009Design}%
  \BibitemOpen
  \bibfield  {author} {\bibinfo {author} {\bibfnamefont {X.}~\bibnamefont
  {Chen}}\ and\ \bibinfo {author} {\bibfnamefont {J.-W.}\ \bibnamefont {Tao}},\
  }\href {\doibase 10.1063/1.3168527} {\bibfield  {journal} {\bibinfo
  {journal} {Appl. Phys. Lett.}\ }\textbf {\bibinfo {volume} {94}},\ \bibinfo
  {pages} {262102} (\bibinfo {year} {2009})}\BibitemShut {NoStop}%
\bibitem [{\citenamefont {Encinosa}\ and\ \citenamefont
  {Etemadi}(1998)}]{Encinosa1998}%
  \BibitemOpen
  \bibfield  {author} {\bibinfo {author} {\bibfnamefont {M.}~\bibnamefont
  {Encinosa}}\ and\ \bibinfo {author} {\bibfnamefont {B.}~\bibnamefont
  {Etemadi}},\ }\href {\doibase 10.1103/PhysRevA.58.77} {\bibfield  {journal}
  {\bibinfo  {journal} {Phys. Rev. A}\ }\textbf {\bibinfo {volume} {58}},\
  \bibinfo {pages} {77} (\bibinfo {year} {1998})}\BibitemShut {NoStop}%
\bibitem [{\citenamefont {Shima}\ \emph {et~al.}(2009)\citenamefont {Shima},
  \citenamefont {Yoshioka},\ and\ \citenamefont {Onoe}}]{Shima2009Geometry}%
  \BibitemOpen
  \bibfield  {author} {\bibinfo {author} {\bibfnamefont {H.}~\bibnamefont
  {Shima}}, \bibinfo {author} {\bibfnamefont {H.}~\bibnamefont {Yoshioka}}, \
  and\ \bibinfo {author} {\bibfnamefont {J.}~\bibnamefont {Onoe}},\ }\href
  {\doibase 10.1103/PhysRevB.79.201401} {\bibfield  {journal} {\bibinfo
  {journal} {Phys. Rev. B}\ }\textbf {\bibinfo {volume} {79}},\ \bibinfo
  {pages} {201401} (\bibinfo {year} {2009})}\BibitemShut {NoStop}%
\bibitem [{\citenamefont {Jensen}\ and\ \citenamefont
  {Dandoloff}(2009)}]{Jensen2009}%
  \BibitemOpen
  \bibfield  {author} {\bibinfo {author} {\bibfnamefont {B.}~\bibnamefont
  {Jensen}}\ and\ \bibinfo {author} {\bibfnamefont {R.}~\bibnamefont
  {Dandoloff}},\ }\href {\doibase 10.1103/PhysRevA.80.052109} {\bibfield
  {journal} {\bibinfo  {journal} {Phys. Rev. A}\ }\textbf {\bibinfo {volume}
  {80}},\ \bibinfo {pages} {052109} (\bibinfo {year} {2009})}\BibitemShut
  {NoStop}%
\bibitem [{\citenamefont {Grivickas}\ \emph {et~al.}(2019)\citenamefont
  {Grivickas}, \citenamefont {Geisz},\ and\ \citenamefont
  {Gupta}}]{Grivickas2019}%
  \BibitemOpen
  \bibfield  {author} {\bibinfo {author} {\bibfnamefont {P.}~\bibnamefont
  {Grivickas}}, \bibinfo {author} {\bibfnamefont {J.~F.}\ \bibnamefont
  {Geisz}}, \ and\ \bibinfo {author} {\bibfnamefont {Y.~M.}\ \bibnamefont
  {Gupta}},\ }\href {\doibase 10.1063/1.5038723} {\bibfield  {journal}
  {\bibinfo  {journal} {Appl. Phys. Lett.}\ }\textbf {\bibinfo {volume}
  {113}},\ \bibinfo {pages} {072101} (\bibinfo {year} {2019})}\BibitemShut
  {NoStop}%
\bibitem [{\citenamefont {Matsutani}\ and\ \citenamefont
  {Tsuru}(1991)}]{doi:10.1143/JPSJ.60.3640}%
  \BibitemOpen
  \bibfield  {author} {\bibinfo {author} {\bibfnamefont {S.}~\bibnamefont
  {Matsutani}}\ and\ \bibinfo {author} {\bibfnamefont {H.}~\bibnamefont
  {Tsuru}},\ }\href {\doibase 10.1143/JPSJ.60.3640} {\bibfield  {journal}
  {\bibinfo  {journal} {J. Phys. Soc. Jpn.}\ }\textbf {\bibinfo {volume}
  {60}},\ \bibinfo {pages} {3640} (\bibinfo {year} {1991})}\BibitemShut
  {NoStop}%
\bibitem [{\citenamefont {Clark}\ and\ \citenamefont
  {Bracken}(1996)}]{CLARK1996}%
  \BibitemOpen
  \bibfield  {author} {\bibinfo {author} {\bibfnamefont {I.~J.}\ \bibnamefont
  {Clark}}\ and\ \bibinfo {author} {\bibfnamefont {A.~J.}\ \bibnamefont
  {Bracken}},\ }\href {http://dx.doi.org/10.1088/0305-4470/29/15/022}
  {\bibfield  {journal} {\bibinfo  {journal} {J. Phys. A: Math. Gen.}\ }\textbf
  {\bibinfo {volume} {29}},\ \bibinfo {pages} {4527} (\bibinfo {year}
  {1996})}\BibitemShut {NoStop}%
\bibitem [{\citenamefont {Zhang}\ \emph {et~al.}(2007)\citenamefont {Zhang},
  \citenamefont {Zhang},\ and\ \citenamefont {Wang}}]{Zhang2007Quantum}%
  \BibitemOpen
  \bibfield  {author} {\bibinfo {author} {\bibfnamefont {E.}~\bibnamefont
  {Zhang}}, \bibinfo {author} {\bibfnamefont {S.}~\bibnamefont {Zhang}}, \ and\
  \bibinfo {author} {\bibfnamefont {Q.}~\bibnamefont {Wang}},\ }\href {\doibase
  10.1103/PhysRevB.75.085308} {\bibfield  {journal} {\bibinfo  {journal} {Phys.
  Rev. B}\ }\textbf {\bibinfo {volume} {75}},\ \bibinfo {pages} {085308}
  (\bibinfo {year} {2007})}\BibitemShut {NoStop}%
\bibitem [{\citenamefont {Schindler}\ \emph {et~al.}(2019)\citenamefont
  {Schindler}, \citenamefont {Wiegand}, \citenamefont {Sichau}, \citenamefont
  {Tiemann}, \citenamefont {Nielsch}, \citenamefont {Zierold},\ and\
  \citenamefont {Blick}}]{Schindler2019}%
  \BibitemOpen
  \bibfield  {author} {\bibinfo {author} {\bibfnamefont {C.}~\bibnamefont
  {Schindler}}, \bibinfo {author} {\bibfnamefont {C.}~\bibnamefont {Wiegand}},
  \bibinfo {author} {\bibfnamefont {J.}~\bibnamefont {Sichau}}, \bibinfo
  {author} {\bibfnamefont {L.}~\bibnamefont {Tiemann}}, \bibinfo {author}
  {\bibfnamefont {K.}~\bibnamefont {Nielsch}}, \bibinfo {author} {\bibfnamefont
  {R.}~\bibnamefont {Zierold}}, \ and\ \bibinfo {author} {\bibfnamefont
  {R.~H.}\ \bibnamefont {Blick}},\ }\href {\doibase 10.1063/1.5001929}
  {\bibfield  {journal} {\bibinfo  {journal} {Appl. Phys. Lett.}\ }\textbf
  {\bibinfo {volume} {111}},\ \bibinfo {pages} {171601} (\bibinfo {year}
  {2019})}\BibitemShut {NoStop}%
\bibitem [{\citenamefont {Duan}\ \emph {et~al.}(2003)\citenamefont {Duan},
  \citenamefont {Niu}, \citenamefont {Sahi}, \citenamefont {Chen},
  \citenamefont {Parce}, \citenamefont {Empedocles},\ and\ \citenamefont
  {Goldman}}]{Duan2003}%
  \BibitemOpen
  \bibfield  {author} {\bibinfo {author} {\bibfnamefont {X.}~\bibnamefont
  {Duan}}, \bibinfo {author} {\bibfnamefont {C.}~\bibnamefont {Niu}}, \bibinfo
  {author} {\bibfnamefont {V.}~\bibnamefont {Sahi}}, \bibinfo {author}
  {\bibfnamefont {J.}~\bibnamefont {Chen}}, \bibinfo {author} {\bibfnamefont
  {J.~W.}\ \bibnamefont {Parce}}, \bibinfo {author} {\bibfnamefont
  {S.}~\bibnamefont {Empedocles}}, \ and\ \bibinfo {author} {\bibfnamefont
  {J.~L.}\ \bibnamefont {Goldman}},\ }\href
  {https://doi.org/10.1038/nature01996} {\bibfield  {journal} {\bibinfo
  {journal} {Nature}\ }\textbf {\bibinfo {volume} {425}},\ \bibinfo {pages}
  {274} (\bibinfo {year} {2003})}\BibitemShut {NoStop}%
\bibitem [{\citenamefont {Jensen}\ and\ \citenamefont
  {Koppe}(1971)}]{Jensen1971Quantum}%
  \BibitemOpen
  \bibfield  {author} {\bibinfo {author} {\bibfnamefont {H.}~\bibnamefont
  {Jensen}}\ and\ \bibinfo {author} {\bibfnamefont {H.}~\bibnamefont {Koppe}},\
  }\href {https://linkinghub.elsevier.com/retrieve/pii/0003491671900315}
  {\bibfield  {journal} {\bibinfo  {journal} {Ann. Phys.}\ }\textbf {\bibinfo
  {volume} {63}},\ \bibinfo {pages} {586} (\bibinfo {year} {1971})}\BibitemShut
  {NoStop}%
\bibitem [{\citenamefont {Wang}\ \emph {et~al.}(2017)\citenamefont {Wang},
  \citenamefont {Jiang},\ and\ \citenamefont {Zong}}]{Wang2017Geometric}%
  \BibitemOpen
  \bibfield  {author} {\bibinfo {author} {\bibfnamefont {Y.-L.}\ \bibnamefont
  {Wang}}, \bibinfo {author} {\bibfnamefont {H.}~\bibnamefont {Jiang}}, \ and\
  \bibinfo {author} {\bibfnamefont {H.-S.}\ \bibnamefont {Zong}},\ }\href
  {\doibase 10.1103/PhysRevA.96.022116} {\bibfield  {journal} {\bibinfo
  {journal} {Phys. Rev. A}\ }\textbf {\bibinfo {volume} {96}},\ \bibinfo
  {pages} {022116} (\bibinfo {year} {2017})}\BibitemShut {NoStop}%
\bibitem [{\citenamefont {Ferrari}\ and\ \citenamefont
  {Cuoghi}(2008)}]{Ferrari2008}%
  \BibitemOpen
  \bibfield  {author} {\bibinfo {author} {\bibfnamefont {G.}~\bibnamefont
  {Ferrari}}\ and\ \bibinfo {author} {\bibfnamefont {G.}~\bibnamefont
  {Cuoghi}},\ }\href {\doibase 10.1103/PhysRevLett.100.230403} {\bibfield
  {journal} {\bibinfo  {journal} {Phys. Rev. Lett.}\ }\textbf {\bibinfo
  {volume} {100}},\ \bibinfo {pages} {230403} (\bibinfo {year}
  {2008})}\BibitemShut {NoStop}%
\bibitem [{\citenamefont {Wang}\ \emph {et~al.}(2016)\citenamefont {Wang},
  \citenamefont {Liang}, \citenamefont {Jiang}, \citenamefont {Lu},\ and\
  \citenamefont {Zong}}]{Wang2016Transmission}%
  \BibitemOpen
  \bibfield  {author} {\bibinfo {author} {\bibfnamefont {Y.-L.}\ \bibnamefont
  {Wang}}, \bibinfo {author} {\bibfnamefont {G.-H.}\ \bibnamefont {Liang}},
  \bibinfo {author} {\bibfnamefont {H.}~\bibnamefont {Jiang}}, \bibinfo
  {author} {\bibfnamefont {W.-T.}\ \bibnamefont {Lu}}, \ and\ \bibinfo {author}
  {\bibfnamefont {H.-S.}\ \bibnamefont {Zong}},\ }\href
  {http://stacks.iop.org/0022-3727/49/i=29/a=295103} {\bibfield  {journal}
  {\bibinfo  {journal} {J. Phys. D: Appl. Phys.}\ }\textbf {\bibinfo {volume}
  {49}},\ \bibinfo {pages} {295103} (\bibinfo {year} {2016})}\BibitemShut
  {NoStop}%
\bibitem [{\citenamefont {Cao}\ \emph {et~al.}(2019)\citenamefont {Cao},
  \citenamefont {Wang}, \citenamefont {Chen}, \citenamefont {Jiang},
  \citenamefont {Xu},\ and\ \citenamefont {Zong}}]{Cao2019}%
  \BibitemOpen
  \bibfield  {author} {\bibinfo {author} {\bibfnamefont {W.-R.}\ \bibnamefont
  {Cao}}, \bibinfo {author} {\bibfnamefont {Y.-L.}\ \bibnamefont {Wang}},
  \bibinfo {author} {\bibfnamefont {X.-L.}\ \bibnamefont {Chen}}, \bibinfo
  {author} {\bibfnamefont {H.}~\bibnamefont {Jiang}}, \bibinfo {author}
  {\bibfnamefont {C.-T.}\ \bibnamefont {Xu}}, \ and\ \bibinfo {author}
  {\bibfnamefont {H.-S.}\ \bibnamefont {Zong}},\ }\href
  {http://www.sciencedirect.com/science/article/pii/S0375960119303160}
  {\bibfield  {journal} {\bibinfo  {journal} {Phys. Lett. A}\ }\textbf
  {\bibinfo {volume} {383}},\ \bibinfo {pages} {2124} (\bibinfo {year}
  {2019})}\BibitemShut {NoStop}%
\end{thebibliography}%

\end{document}